\documentclass[%
reprint,
superscriptaddress,
 amsmath,amssymb,
 aps,
prd,
]{revtex4-2}

\usepackage{graphicx}
\usepackage{dcolumn}
\usepackage{bm}

\usepackage{xcolor} 

\begin{document}

\title{SATDI: Simulation and Analysis for Time-Delay Interferometry}

\author{Gang Wang}
\email[Gang Wang: ]{gwang@shao.ac.cn, gwanggw@gmail.com}
\affiliation{Shanghai Astronomical Observatory, Chinese Academy of Sciences, Shanghai 200030, China}

\date{\today}

\begin{abstract}

Time-delay interferometry (TDI) is essential for space-based gravitational wave (GW) missions to effectively suppress laser frequency noise and achieve targeting sensitivity. The principle of the TDI is to synthesize multiple laser link measurements between spacecraft and create virtual equal-arm interferometry. This process blends instrumental noises and tunes the response function to GW, yielding data characterized by TDI combinations. Extracting signals requires modeling GW signals under TDI operations in the frequency domain.
In this work, we introduce a versatile framework, SATDI, which integrates simulation and analysis for TDI. The simulation aims to implement TDI to instrumental noises and GW signals, investigate influential factors in noise suppressions, and explore GW characterizations across different TDI configurations.
The analysis component focuses on developing robust algorithms for modeling TDI responses to extract GWs and accurately determine source parameters. LISA is selected as the representative space mission to demonstrate the effectiveness of our framework. We simulate and analyze data containing GW signals from massive black hole binary coalescence, examining data from both first-generation and second-generation TDI Michelson configurations. The results not only validate the framework but also illustrate the influence of different factors on parameter estimation.

\end{abstract}


\maketitle

\section{Introduction}

LISA is planned to be launched in the mid 2030s and open a new era of gravitational wave (GW) observation in the milli-Hz band \cite{2017arXiv170200786A,Colpi:2024xhw}. The mission will feature a triangle constellation formed by three spacecraft (S/C). Drag-free technology will be employed to maintain test-masses housed within each S/C on geodesic trajectories. Laser interferometry will be utilized to precisely monitor the relative motion between test-mass caused by GWs. However, due to orbital dynamics and laser stability, the laser frequency noise will be orders of magnitude higher than the desired sensitivity for GW observations. To overcome the laser noise, time delay interferometry (TDI) was developed for GW space missions \cite{1997SPIE.3116..105N,1999ApJ...527..814A}. Based on the laser noise suppression capabilities of TDI, two generations have been categorized. The first-generation TDI configurations could cancel laser frequency fluctuations in a static unequal arm constellation \cite[and references therein]{1999ApJ...527..814A,2000PhRvD..62d2002E,2001CQGra..18.4059A,Larson:2002xr,Dhurandhar:2002zcl,2003PhRvD..67l2003T,Vallisneri:2004bn,2008PhRvD..77b3002P,Tinto:2020fcc}, while the second-generation could further suppress the laser noise in time-varying arm lengths \cite[and references therein]{Shaddock:2003dj,Cornish:2003tz,Tinto:2003vj,Vallisneri:2005ji,Dhurandhar:2010pd,Tinto:2018kij,2019PhRvD..99h4023B,2020arXiv200111221M,Vallisneri:2020otf}.

To assess the feasibility and capabilities of TDI, several simulators have been developed for the LISA mission. \textsf{LISA Simulator} was built to simulate the data and response function in the frequency domain \cite{Cornish:2002rt,Rubbo:2003ap}. \textsf{Synthetic LISA} was developed to investigate scientific and technical requirements for original LISA \cite{Vallisneri:2004bn}. \textsf{TDISim} was developed to study the technical solutions and data simulation of TDI \cite{Otto:2015}. LISACode \cite{Petiteau:2008zz} and the successor LISANode \cite{LISANode} aim to perform more specific and systematic TDI simulation for LISA, incorporating functional components such as \textsf{PyTDI} \cite{PyTDI}, \textsf{LISA GW Response} \cite{LISAresponse}, and \textsf{LISA instrument} \cite{LISAinstrument}. On the other side, experimental demonstrations have also (partially) validated the feasibility of TDI \cite{deVine:2010vf,Vinckier:2020qlu}.

Another essential aspect of TDI studies is the extraction of signals from TDI data streams, where the GWs are characterized by the arrangement of TDI link combinations. Determining source parameters requires modeling the TDI response in the frequency domain. Additionally, as the primary targeting source of space missions, GW signals from massive black hole binaries (MBHB) could persist for days to months in the sensitive frequency band, and the antenna patterns with orbital motions must also be taken into account.
With the modeled GW signals under TDI, classical algorithms for inferring parameters of MBHBs include Markov chain Monte Carlo (MCMC) or nested sampler \cite{Cornish:2020vtw,Marsat:2020rtl,Katz:2021uax,Pratten:2022kug,Karnesis:2023ras}. To expedite likelihood calculations, GPU acceleration and heterodyne (or relative binning) algorithms have been developed \cite{Katz:2020hku,Cornish:2021lje,Zackay:2018qdy}. Specifically focusing on the MBHBs, \textsf{BBHx} was developed to enhance LISA data analysis for massive binaries with GPU acceleration \cite{Katz:2020hku,Katz:2021uax}, and PyCBC has extended chirp signal identification and inference from ground-based to space-borne interferometers \cite{Weaving:2023fji}. Moreover, machine learning techniques have been applied to boost GW parameter inference \cite[and reference therein]{Williams:2021qyt,Gabbard:2019rde,Langendorff:2022fzq,Dax:2022pxd,Ruan:2021fxq,Ruan:2023fce,Du:2023plr,Sun:2023vlq}. 
On the other side, unlike ground-based GW observations, space-based detectors simultaneously detect GWs from various sources. A global analysis method has been proposed to identify and infer the parameters of sources simultaneously \cite{Littenberg:2020bxy,Littenberg:2023xpl}. \textsf{Eryn}, as a multipurpose toolbox for Bayesian inference, has been developed to fulfill the global fitting algorithm \cite{Karnesis:2023ras}.

In this work, we present a generic framework, SATDI, which integrates simulation and analysis of TDI. The simulation aims to 1) implement various TDI combinations for instrumental noises and GW signals in both the time-domain and frequency-domain, 2) evaluate TDI noise levels and diagnose the influential factors, and 3) characterize GW signals across different TDI configurations. The analysis focuses on developing robust algorithms to identify GWs and accurately determine the source parameter in the frequency-domain. Furthermore, the framework will explore new TDI observables and evaluate their performance in noise suppressions and GW analysis. In a companion paper \citet{Wang:2024b}, we utilize this framework to introduce a novel TDI configuration referred as hybrid Relay and validate its capabilities in laser noise suppression, clock noise cancellation, and the robustness in inferring the GW signal from MBHBs. In current study, employing LISA as the representative case, we outline the procedures and algorithms for simulation and analysis. Data containing a GW signal from coalescing MBHB are simulated for both first-generation and second-generation TDI Michelson configurations, and the analysis is performed with different data combinations and influential factors. The results not only confirm the validity of the framework but also illustrate the influence of different factors on parameter estimation. 

This paper is organized as follows:
In Section \ref{sec:algorithm}, we introduce the numerical algorithm to pre-calculate the time delays based on the geometry of TDI. The simulation processes are presented in Section \ref{sec:simulation}, where we specify the noise budgets and formulate the GW response in TDI.
In Section \ref{sec:analysis}, we demonstrate parameter inferences with different TDI observables and setups.
A brief conclusion and discussion are given in Sec. \ref{sec:conclusions}.
(We set $G=c=1$ in this work unless otherwise specified in the equations.)

\section{Time delay interferometry} \label{sec:algorithm}

\subsection{Simulation of mission orbits}

LISA is designed to employ three S/C to form a triangular constellation with an arm length of $2.5 \times 10^6$ km, trailing the Earth by 20$^\circ$. TAIJI is another space GW mission utilizing a LISA-like orbit with a triangular interferometer of $3 \times 10^6$ km, leading/trailing the Earth by 20$^\circ$ \cite{Hu:2017mde,Ruan:2020smc,Wang:2020a,Wang:2021uih,Wang:2021njt}. These two orbital formations could be formulated using the Clohessy-Wiltshire frame \cite{Dhurandhar:2004rv,Nayak:2006zm}. In previous works, we have developed a workflow to optimize the LISA-like orbits using an ephemeris framework and calculate the mismatches of TDI combinations in dynamic cases \cite{Wang:2012ce,Dhurandhar:2011ik,Wang:2017aqq}. In addition to LISA and TAIJI, the orbits of ASTROD-GW \cite{Ni:2011ib,Ni:2012eh} and AMIGO are calculated for free-falling trajectories \cite{Wang:2011,Wang:2014aea,Wang:2014cla,Wang:2012te,Ni:2019nau}. All these missions are proposed to utilize TDI and their numerical orbits serve as input for our framework.

The ephemeris framework serves as a fundamental tool for orbital simulation. To calculate the orbit with sufficient accuracy, various interactions are considered, including 1) Newtonian and first-order post-Newtonian interactions among the Sun, major planets, Pluto, Moon, Ceres, Pallas, and Vesta; 2) figure interactions between the Sun, Earth, Moon, and other bodies treated as point masses; 3) perturbations from a selected set of 340 asteroids; and 4) tidal effects on the Moon caused by the Earth \cite{DE430,2021AJ....161..105P}. Once the initial conditions of the S/C and celestial bodies are set at starting mission time, the numerical integration is carried out under the gravitational forces in the solar system. The coordinate time of orbit calculations is based on barycentric dynamical time (TDB), and the time difference between the proper time on the S/C and coordinate time could be calculated by including the relativistic effects as specified in \cite{Soffel:2003cr}. 

These pre-calculated orbit data are essential components of simulation and analysis. While simulation can utilize the orbital data directly assuming the positions and velocities are true values for the missions, the analysis should be based on the observed orbital data with errors in principle. The orbit positioning for a deep space mission with a separation of $\sim 5 \times 10^8$ km from Earth may have an uncertainty ranging from less than a kilometer to tens of kilometers depending on ground tracking data and inter-S/C measurements \cite{YANG20221060,Li:2024}. Our tests show that this level of orbital positioning error has a trivial impact on the current analysis. Therefore, we temporarily ignore the errors of orbit determination during the analysis process.
In this work, we select the LISA orbit previously obtained in \cite{Wang:2017aqq}, which meets the requirement of relative velocities between S/C less than 5 m/s for $2.5 \times 10^6$ km arm length \cite{2017arXiv170200786A}. 

\subsection{Formulation and geometry of TDI} \label{subsec:spacecraft}

\begin{figure}[htb]
\includegraphics[width=0.48\textwidth]{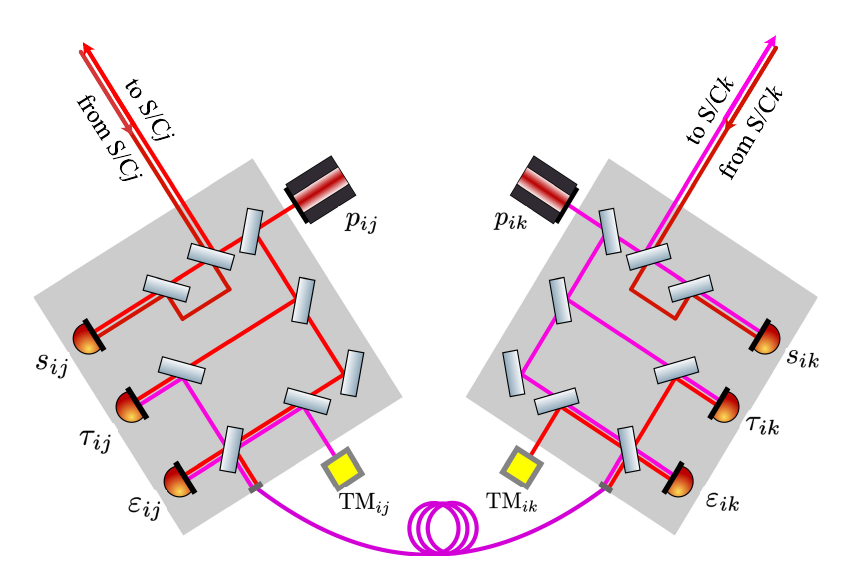} 
\caption{\label{fig:SC_layout} The measurement designs on S/C$i$ of LISA mission \cite{Otto:2015}.}
\end{figure}

TDI constructs equivalent equal arm interferometry by combining multiple link measurements. In the current payload designs of LISA, two optical benches are deployed on each S/C, and the three measurements are gathered on each optical bench \cite[and references therein]{Otto:2012dk,Otto:2015,Tinto:2018kij,2019PhRvD..99h4023B}. The interferometric measurements on S/C$i$ are illustrated in Fig. \ref{fig:SC_layout}.

In Fig. \ref{fig:SC_layout}, the $s_{ji}$, $\varepsilon_{ij}$, and $\tau_{ij}$ refer to the science interferometer, test-mass interferometer, and reference interferometer on each optical bench, respectively. With only considering laser noise, acceleration noise, and optical metrology noise, three interferometric measurements are expressed as 
\begin{equation} \label{eq:s_epsilon_tau}
\begin{aligned}
   s_{ij} = & \mathcal{D}_{ji} p_{ji}(t) - p_{ij}(t) + n^{\rm op}_{ij}(t), \\
   \varepsilon_{ij} = & p_{ik}(t) - p_{ij}(t) + 2 n^{\rm acc}_{ij}(t) , \\
   \tau_{ij} = & p_{ik}(t) - p_{ij}(t) , 
\end{aligned}
\end{equation}
where $p_{ij}$ denotes laser noise on the optical bench of S/C$i$ pointing to S/C$j$, $n^{\mathrm{op}}_{ij}$ represents optical path noise on the S/C$i$ facing $j$, $n^{\mathrm{acc}}_{ij}$ denotes the acceleration noise from test mass on the S/C$i$ pointing to $j$, and $\mathcal{D}_{ij}$ is a time-delay operator, $\mathcal{D}_{ij}y = y(t- L_{ij})$ where $L_{ij}$ is the arm length or propagation time from S/C$i$ to $j$. 

\begin{figure}[htb]
\includegraphics[width=0.23\textwidth]{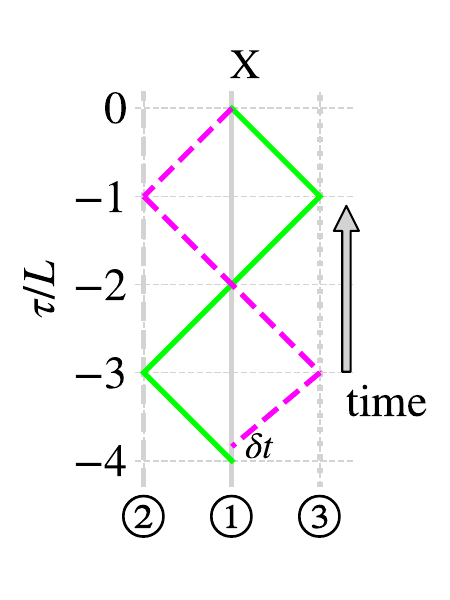} 
\includegraphics[width=0.24\textwidth]{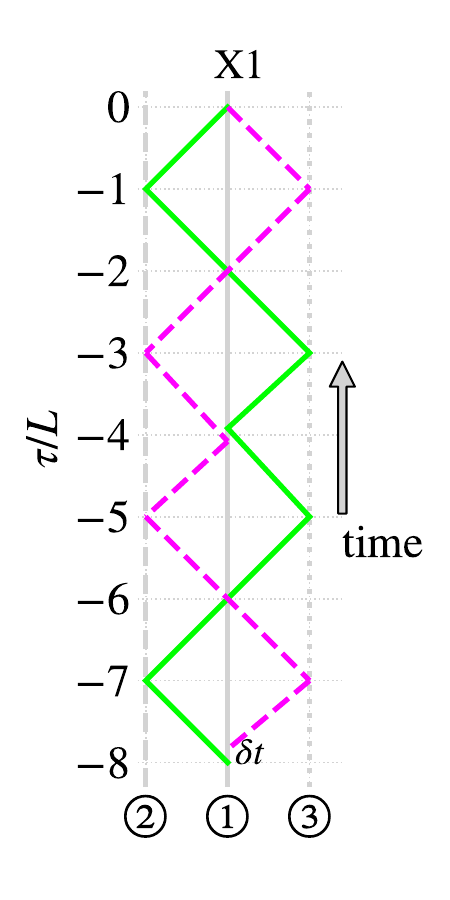} 
\caption{\label{fig:X_X1_diagram} The geometric diagrams for the TDI channels of first-generation Michelson-X (left) and second-generation X1 (right). The vertical lines indicate the trajectories of three S/C (\textcircled{$i$} indicates S/C$i$, $i = 1, 2, 3$), and the horizontal separations between vertical lines indicate the spatial distances between S/C. The green and magenta lines represent two interferometric beams in the TDI link combination.}
\end{figure}

The expressions of TDI have been formulated in previous literature \cite{1999ApJ...527..814A,2000PhRvD..62d2002E,2001CQGra..18.4059A,Larson:2002xr,2003PhRvD..67l2003T,Tinto:2003vj,Shaddock:2003dj,Tinto:2020fcc}. As the fiducial TDI configuration, the expression of the first-generation Michelson-X could be written as
\begin{equation} \label{eq:X_measurement}
\begin{aligned}
{\rm X} =& [ \eta_{31} + \mathcal{D}_{31} \eta_{13}  + \mathcal{D}_{31} \mathcal{D}_{13}  \eta_{21} + \mathcal{D}_{31} \mathcal{D}_{13} \mathcal{D}_{21} \eta_{12} ] \\
 & -  [ \eta_{21} + \mathcal{D}_{21} \eta_{12} + \mathcal{D}_{21} \mathcal{D}_{12} \eta_{31} + \mathcal{D}_{21} \mathcal{D}_{12} \mathcal{D}_{31} \eta_{13} ],
\end{aligned}
\end{equation}
where $\eta_{ji}$ is a combination of interferometric measurements from S/C$j$ to S/C$i$ as defined in \citep{Otto:2012dk,Otto:2015,Tinto:2018kij}. For the links in the counterclockwise directions ($j=$S/C2 $\rightarrow $ $i=$ S/C1, S/C3$\rightarrow$S/C2 and S/C1$\rightarrow $S/C3), the $\eta_{ji}$ are 
\begin{equation} \label{eq:eta1}
\begin{aligned}
  \eta_{ji} &= s_{ji} + \frac{1}{2} \left[ \tau_{ij} - \varepsilon_{ij} + \mathcal{D}_{ji} ( \tau_{ji} - \varepsilon_{ji} ) + \mathcal{D}_{ji} ( \tau_{ji} - \tau_{jk} ) \right].
\end{aligned}
\end{equation}
For the links in the clockwise directions, their expressions are slightly different,
\begin{equation} \label{eq:eta2}
\begin{aligned}
  \eta_{ji} &= s_{ji} + \frac{1}{2} \left[ \tau_{ij} - \varepsilon_{ij}  + \mathcal{D}_{ji} ( \tau_{ji} - \varepsilon_{ji} ) + (\tau_{ik} -  \tau_{ij}) \right].
\end{aligned}
\end{equation}
The corresponding second-generation TDI channels Michelson-X1 could be written as
\begin{equation} \label{eq:X1_measurement}
\begin{aligned}
{\rm X1} \simeq & - \mathrm{X} + \mathcal{D}_{21} \mathcal{D}_{12} \mathcal{D}_{31}  \mathcal{D}_{13} [ \eta_{31} + \mathcal{D}_{31} \eta_{13}  + \mathcal{D}_{31} \mathcal{D}_{13} \eta_{21} \\ 
&+ \mathcal{D}_{31} \mathcal{D}_{13} \mathcal{D}_{21} \eta_{12}  ] + \mathcal{D}_{31} \mathcal{D}_{13} \mathcal{D}_{21}  \mathcal{D}_{12} [ \eta_{21} + \mathcal{D}_{21} \eta_{12} \\ 
& + \mathcal{D}_{21} \mathcal{D}_{12} \eta_{31} + \mathcal{D}_{21} \mathcal{D}_{12} \mathcal{D}_{31} \eta_{13}  ].
\end{aligned}
\end{equation}
Their geometries are illustrated in Fig. \ref{fig:X_X1_diagram}, which will guide time-delay calculations in the following subsection. 

For each TDI configuration, whether first-generation or second-generation, three ordinary channels could be constructed by cyclically permuting the indexes of three S/C. Then three (quasi-)orthogonal or optimal observables, denoted as (A, E, T), can be composed from ordinary observables $(a, b, c)$ using a linear combination \cite{Prince:2002hp,Vallisneri:2007xa},
\begin{equation} \label{eq:abc2AET}
\begin{bmatrix}
\mathrm{A}  \\ \mathrm{E}  \\ \mathrm{T} 
\end{bmatrix}
 = 
\begin{bmatrix}
-\frac{1}{\sqrt{2}} & 0 & \frac{1}{\sqrt{2}} \\
\frac{1}{\sqrt{6}} & -\frac{2}{\sqrt{6}} & \frac{1}{\sqrt{6}} \\
\frac{1}{\sqrt{3}} & \frac{1}{\sqrt{3}} & \frac{1}{\sqrt{3}}
\end{bmatrix}
\begin{bmatrix}
a \\ b  \\ c 
\end{bmatrix}.
\end{equation}
The A and E channels are expected to be science data streams capable of effectively detecting GW, while the T channel is expected to be GW-insensitive data and to characterize instrumental noises.

\subsection{Algorithm for time delay calculation} \label{subsec:algorithm}

Time delays are key quantities in TDI used to operate interferometric measurements as shown in Eqs. \eqref{eq:s_epsilon_tau}-\eqref{eq:X1_measurement}.
Their calculations are implemented along the geometry of a TDI combination as illustrated in Fig. \ref{fig:X_X1_diagram}. Three vertical lines in Fig. \ref{fig:X_X1_diagram} represent the trajectories of three S/C in the time direction. The horizontal separations between vertical lines indicate the spatial distances between S/C. The green and magenta lines show the two beams in the TDI link combination. The ticks on the y-axis represent the time delay with respect to the interferometric time $\tau=0$. 
We select the Michelson-X observable to briefly explain the procedures of calculation, and a more detailed description can be found in \cite{Wang:2020pkk}. The calculation starts from S/C1 at the earliest time point $ \tau \simeq -4L$ moving toward S/C2 at $\tau \simeq -3L$ along the green line. The propagation time and position of laser arrival are determined at S/C2. The second step continues from S/C2 to S/C1 along the path of the green line to obtain the laser travel time and position of S/C1. When the calculation along green line reaches the S/C1 at the latest time point $\tau = 0$, the calculation continues along the magenta line in the backward time direction. The calculation stops at the initial sending S/C after completing a loop. Afterward, the time delays with respect to the latest time points are resolved from the result from each step, as well as the S/C positions. 

During the propagation time calculation, additional relativistic time delays caused by gravitational fields are considered \cite{Ashby:2008lea}. 
The time delay from the sending time $T^s$ at $\bm{r}_s$ to the receiving time $T^r$ at $\bm{r}_r$ is obtained by \cite{Shapiro:1964uw,Kopeikin:2008xv},
\begin{equation} \label{eq:L_ij}
 T^{r} - T^{s} =  \frac{R}{c} + \Delta T_{\text{PN}},
\end{equation}
where $R$ is the coordinate distance between the sender and receiver, $c$ is the speed of light, and $\Delta T_{\text{PN}}$ is the relativistic time delay caused by gravitational field,
   \begin{equation}
   \begin{split}
   \Delta T_{\text{PN}} & = \frac{2GM}{c^3} \ln \left( \frac{R_s+R_r+R}{R_s+R_r-R} \right) \\
    + &  \frac{G^2 M^2}{c^5} \frac{R}{R_s R_r} \left[ \frac{15}{4} \frac{\arccos({\bf{N}}_s \cdot {\bf{N}}_r ) }
    {|{\bf{N}}_s \times {\bf{N}}_r|} -\frac{4}{1+ \bf{N}_s \cdot \bf{N}_r } \right],
    \end{split}
   \end{equation}
where $G$ is the gravitational constant, $M$ is the mass of the celestial body, $\bf{N}_s$ and $\bf{N}_r$ are unit vectors from the celestial body to the $\bm{r}_s$ and $\bm{r}_r$ positions, and $R_s$ and $R_r$ represent the radial distances of sender and receiver from the gravitating body, respectively. In the current calculation, the leading order of relativistic time delay caused by the gravitational field of the Sun is included, the effects from other planets are disregarded. Additionally, due to the orbital motion of the S/C, an iteration process is utilized to determine the arrival time and position of the receiver,
 \begin{equation} \label{eq:iteration}
 \begin{aligned}
      T^r_0 &= T^s_0 + T_1 + T_2 + T_3 +...  \\
       T_1      &= \frac{| \bm{r}_r (T^s_0)-\bm{r}_s (T^s_0) |}{c} + \Delta T_{1,\text{PN}}  \\
      T_1 + T_2  &= \frac{| \bm{r}_r (T^s_0 +T_1)-\bm{r}_s (T^s_0) |}{c} + \Delta T_{2,\text{PN}}  \\
     T_1 + T_2 +T_3 &=\frac{| \bm{r}_r (T^s_0+T_1+T_2)-\bm{r}_s(T^s_0) |}{c} + \Delta T_{3,\text{PN}}  \\
   & ......
\end{aligned}
\end{equation}
The Chebyshev polynomial interpolation is employed to precisely determine the position of S/C at any given moment \cite{Newhall:1989CeMec,Li&Tian:2004}.

In this calculation, the propagation time between S/C are computed numerically based on their orbit data. In a realistic scenario, the ranging time from sender to receiver would be estimated from measurements taken on S/C \cite[references therein]{Heinzel:2011zz,Reinhardt:2023ccg,Tinto:2004yw,Page:2023hxm}. 
In the simulation of the next section, the propagation time with a ranging error is taken into account to evaluate the residual laser noise in the TDI channels. For analysis of the GW signal from MBHB in Section \ref{sec:analysis}, the uncertainties of arm lengths are expected to be trivial and negligible.

\section{Simulation of TDI} \label{sec:simulation}

\subsection{Simulation diagram}

\begin{figure}[htb]
\includegraphics[width=0.46\textwidth]{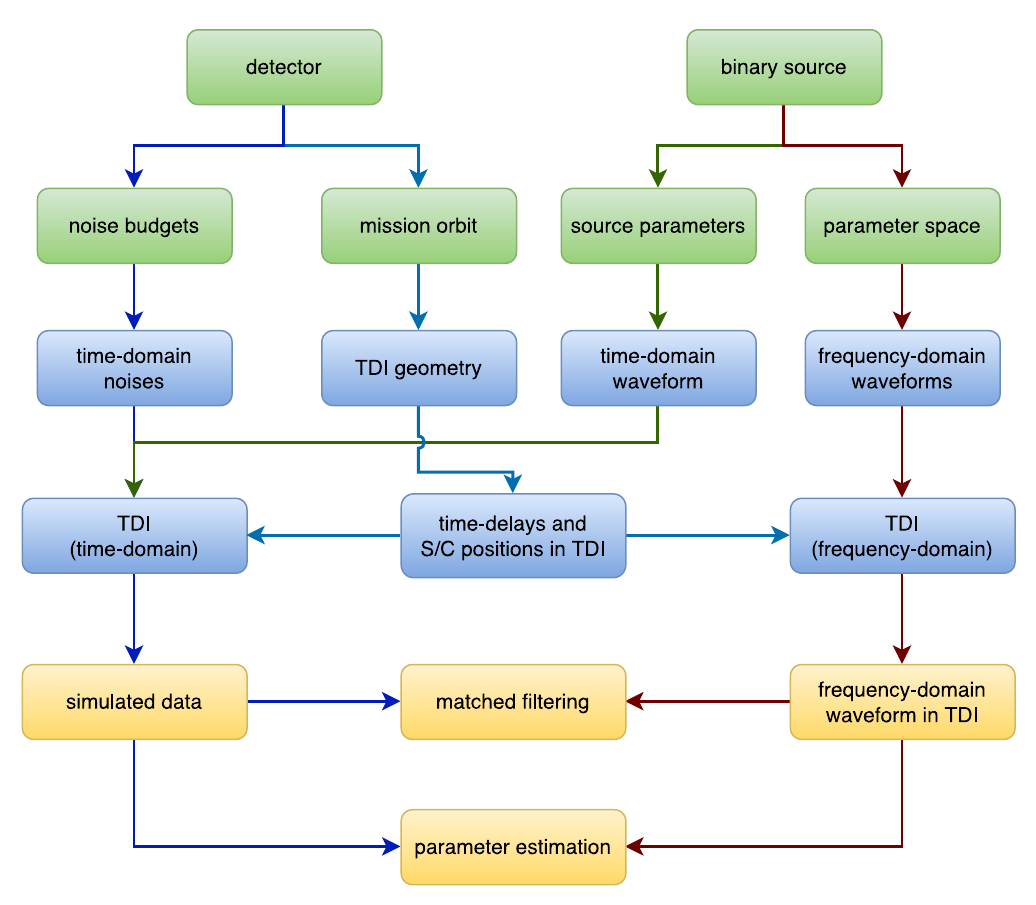}
\caption{\label{fig:diagram} The diagram of the simulation and analysis. The green modules show the setups, the blue modules represent the processes of simulation, and the yellows indicate the analysis content.}
\end{figure}

The diagram illustrating simulation and analysis process is presented in Fig. \ref{fig:diagram}. The initial step involves selecting a mission with noise budgets and setting targeting source(s) is(are) set. In the second step, a TDI configuration is chosen, and the geometry of the link combination is generated accordingly. Subsequently, the time delays and positions of S/C are calculated by implementing algorithms in Section \ref{sec:algorithm}. Based on the noise budgets, time-domain noises are generated for each component with a specified sample rate (In this work, we choose 4 Hz), assuming that the different noises are uncorrelated. TDI is then implemented by synthesizing these noises, and the data at each time point is obtained by using Lagrange interpolation \cite{Shaddock:2004ua,Bayle:2019dfu}. After TDI, the noise data are yielded with laser noise suppression.

On the other hand, a GW signal from MBHB is simulated using a set of chosen parameters. The GW waveform in the time-domain is generated from the waveform model SEOBNRv4HM \cite{Bohe:2016gbl}. Subsequently, TDI is applied to the waveform in the SSB coordinate with TDB, and the responded waveform is computed considering the orbital motion of detector. The waveform is then injected into the noise data and yields the simulated data containing a GW signal. In the next step, the analysis is performed in the frequency domain. The frequency-domain waveforms are generated from a targeted parameter space using the reduced order model (ROM) of SEOBNRv4HM \cite{Cotesta:2020qhw}. The waveforms multiplied by the response function of TDI are used for matched filtering to identify GW signs and infer the source parameters.

\subsection{Noises}

There are various types of observation noises for LISA \cite{Bayle:2022okx}, and our current focus is on laser frequency noise, acceleration noise, and optical metrology noise. The dominant laser frequency noise is generated by Nd:YAG lasers with a stability of $30 \ \mathrm{Hz}/\sqrt{\mathrm{Hz}}$, corresponding to an amplitude spectral density (ASD) of $\tilde{p}(f) \simeq 1 \times 10^{-13} / \sqrt{\rm Hz}$ \cite{2017arXiv170200786A}. In this investigation, the first-generation and second-generation TDI Michelson observables are examined for laser noise suppressions. A ranging error between S/C is considered which includes a bias of 3 ns (0.9 m) and noise with a spectrum of $1 \times 10^{-15} / f / \sqrt{\rm Hz}$ \cite{Bayle:2022okx,Hartwig:2022yqw,Staab:2023qrb}.

Clock noise is assumed to be sufficiently canceled out by using the algorithm in \cite{Hartwig:2020tdu} and is therefore ignored in current simulation. Both the acceleration noise $\mathrm{ S_{n, acc} }$ and optical path noise $\mathrm{S_{n, op}}$ are included with spectra
\begin{equation}
\begin{aligned}
& \sqrt{ S_{\mathrm{acc}, ij} } = A_{\mathrm{acc}, ij} \frac{\rm fm/s^2}{\sqrt{\rm Hz}} \sqrt{1 + \left(\frac{0.4 {\rm mHz}}{f} \right)^2 }  \sqrt{1 + \left(\frac{f}{8 {\rm mHz}} \right)^4 }, \\
& \sqrt{ S_{\mathrm{op}, ij} } = A_{\mathrm{op}, ij} \frac{\rm pm}{\sqrt{\rm Hz}} \sqrt{1 + \left(\frac{2 {\rm mHz}}{f} \right)^4 },
 \end{aligned}
\end{equation}
where $A_{\mathrm{acc},ij}$ and $A_{\mathrm{op}, ij}$ are tunable amplitudes of noise for each optical bench, with $A_{\mathrm{acc}}=3$ and $A_{\mathrm{op}} = 10$ assigned to all components in current setups.

\begin{figure}[htb]
\includegraphics[width=0.48\textwidth]{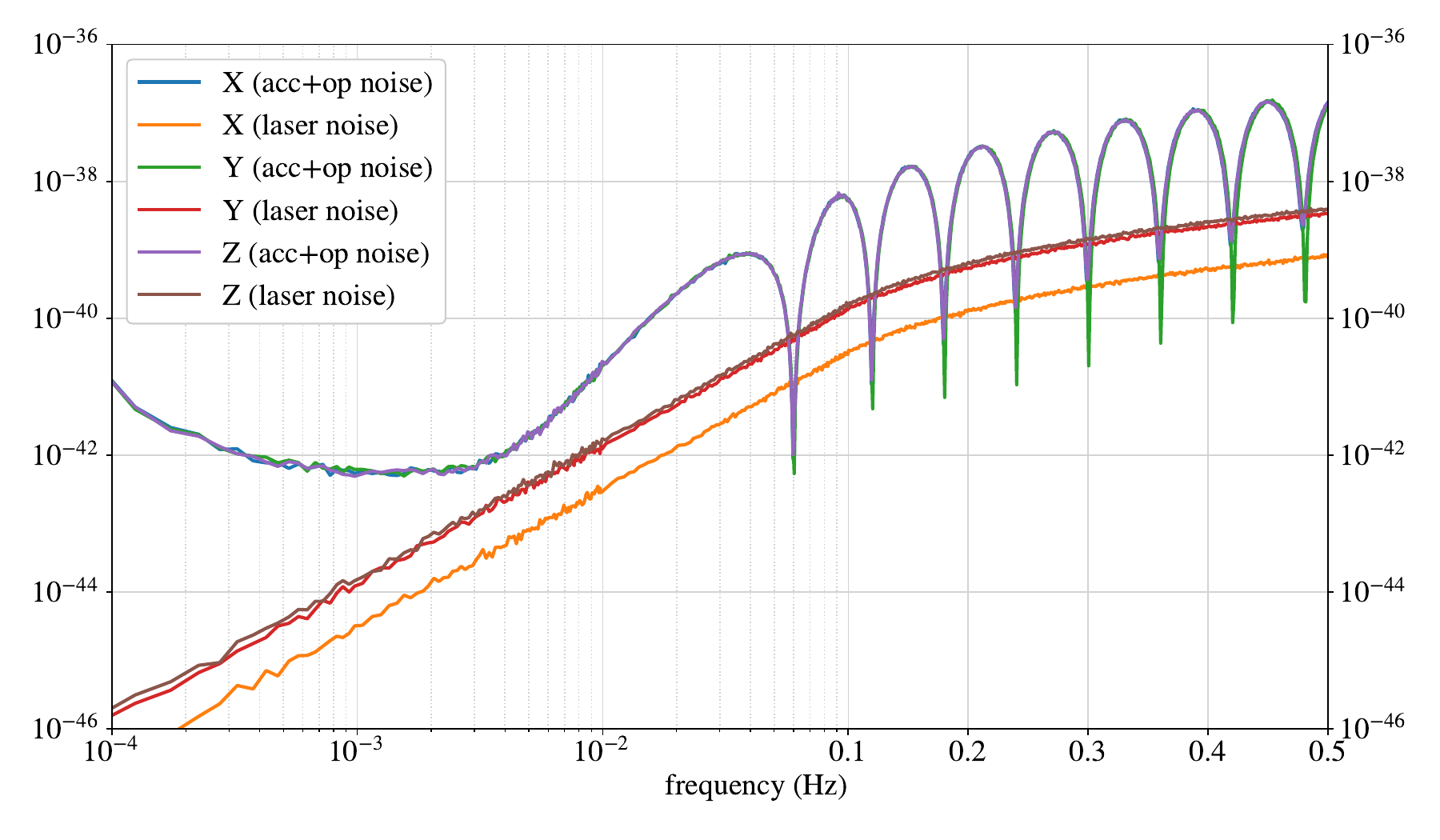}
\includegraphics[width=0.48\textwidth]{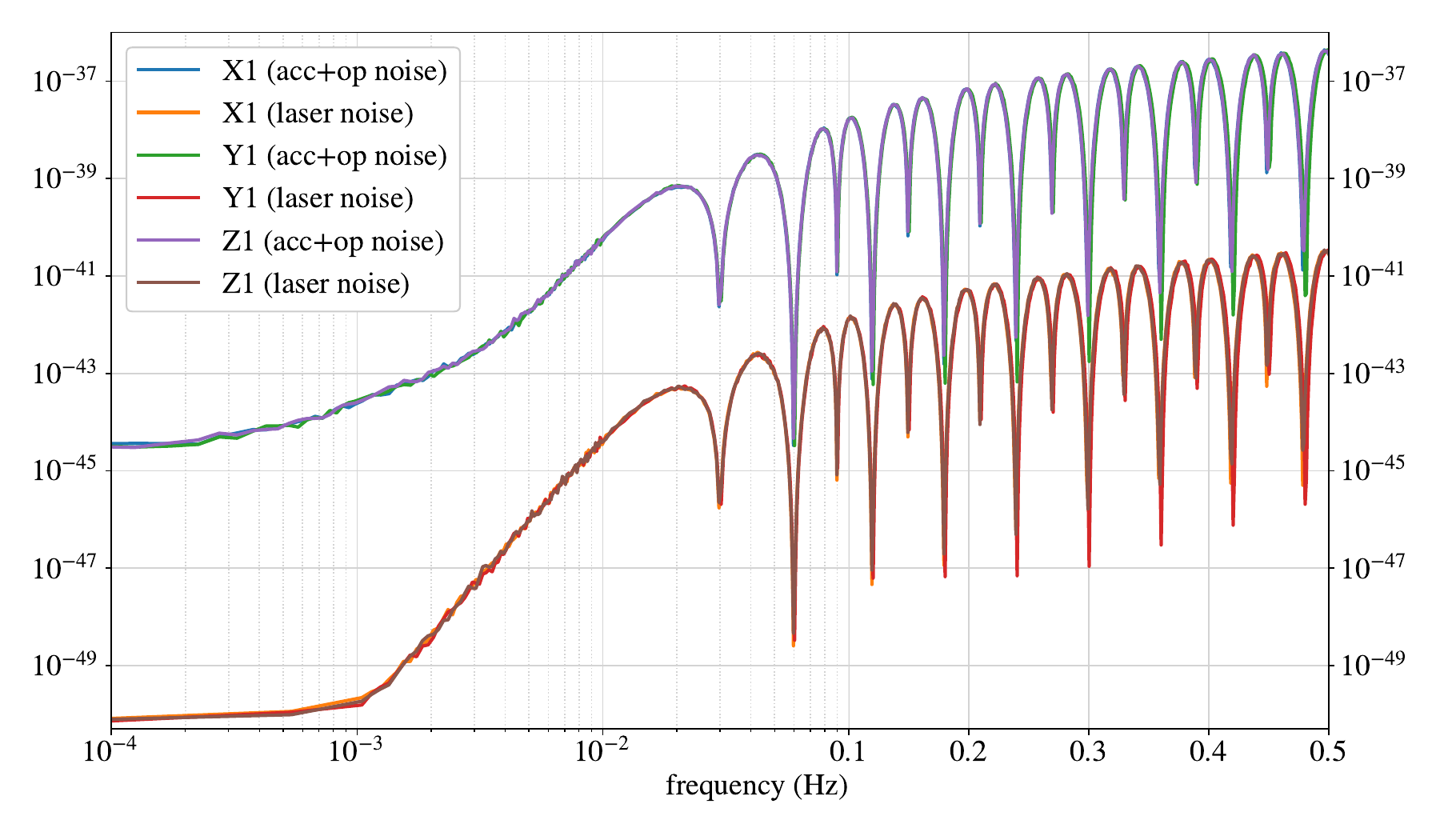}
\caption{\label{fig:psd_simulation} The noise PSDs of the first-generation (upper) and second-generation (lower) TDI Michelson observables. In the first-generation channels, the residual laser noise is mostly lower than the secondary noise (acc+op, acceleration noise + optical metrology noise) in current simulated data, except at the characteristic frequencies. In the second-generation TDI channel, the laser noise is orders of magnitude lower than secondary noises. (Note that a log-scale of x-axis is used for frequencies lower than 0.1 Hz in the x-axis, and a linear scale is utilized for higher frequencies.)}
\end{figure}

The examinations of laser noise suppressions in Michelson configurations are performed as tests, and the PSDs of residual laser noise are presented in Fig. \ref{fig:psd_simulation}. The results for the first-generation TDI channels, (X, Y, Z), are displayed in the upper plot, and the results for the second-generation TDI channels (X1, Y1, Z1) are depicted in the lower panel. Compared to the secondary noises including acceleration noise and optical metrology noise, residual laser noises in first-generation TDI channels are moderately lower at lower frequencies and slightly higher at characteristic frequencies in the current dataset. This level of laser noise is mainly limited by the path mismatches of two beams due to the orbital dynamics. For the second-generation TDI channels, the residual laser noise is significantly lower than secondary noises, and this level of suppression is subject to the ranging errors between S/C. 

\subsection{GW waveforms} \label{subsec:waveform}
 
 \begin{figure}[htb]
\includegraphics[width=0.48\textwidth]{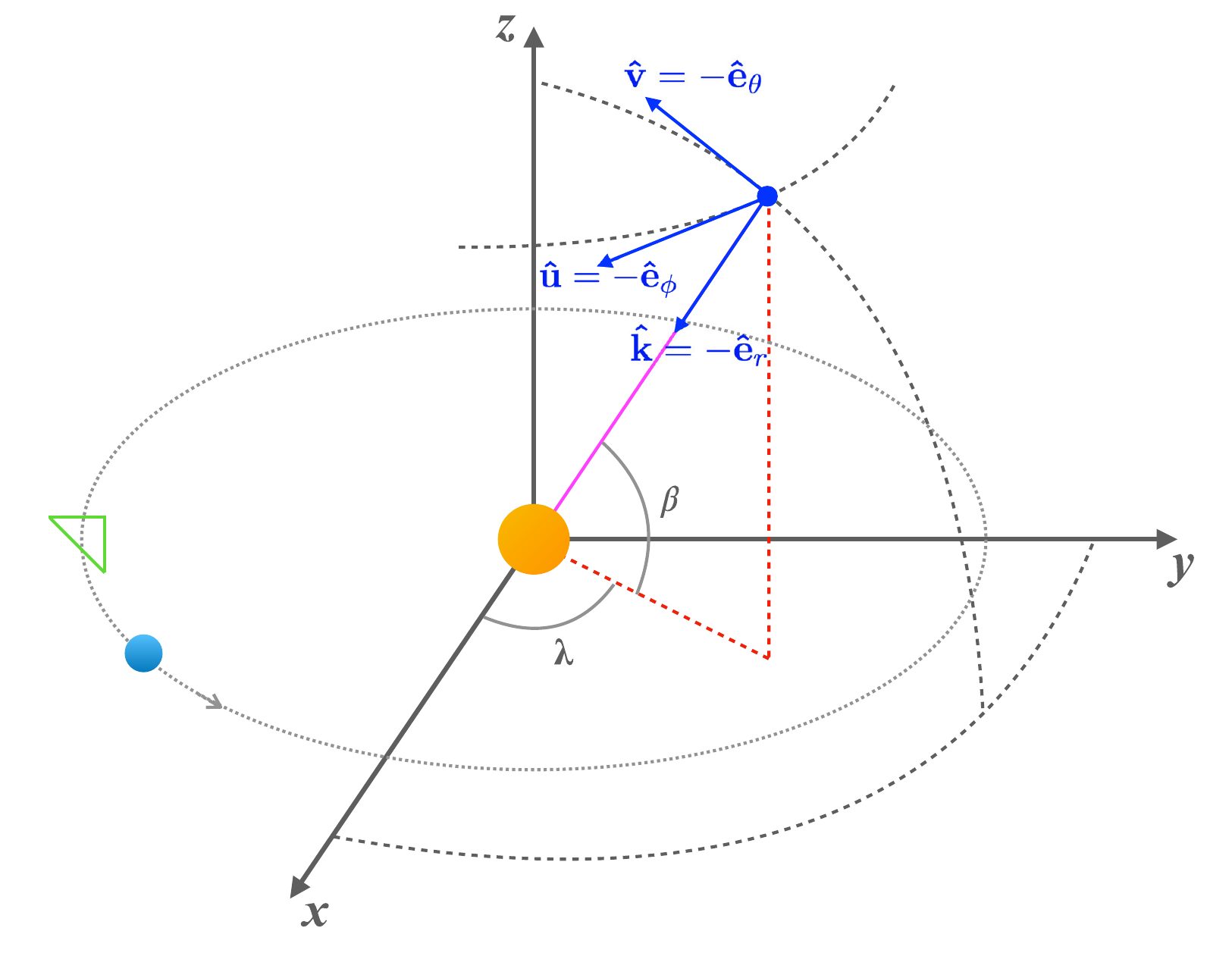} 
\caption{\label{fig:SSB_frame} The conventions in the SSB frame \cite{Katz:2022yqe}.}
\end{figure}

The conventions for waveform simulation in SSB coordinates are illustrated in Fig. \ref{fig:SSB_frame} \cite{Vallisneri:2012np,Katz:2022yqe}. For a source of MBHB, its location is described by the ecliptic longitude $\lambda$ and ecliptic latitude $\beta$. The propagation vector of GW will be 
\begin{equation}
\mathbf{k} = - \mathbf{e}_r = ( -\cos \beta \cos \lambda, -\cos \beta \sin \lambda, -\sin \beta ).
\end{equation}
The orthogonal vectors in the longitude and latitude directions are defined as 
\begin{align}
\mathbf{u} = - \mathbf{e}_\phi = & ( \sin \lambda, - \cos \lambda, 0 ), \\
\mathbf{v} = - \mathbf{e}_\theta = & ( -\sin \beta \cos \lambda, -\sin \beta \sin \lambda, \cos \beta ).
\end{align}
By defining the polarization angle $\psi$, the principal directions $[\mathbf{p}, \mathbf{q}]$ are rotated from $[\mathbf{u}, \mathbf{v}]$, 
\begin{align}
 \mathbf{\epsilon}_1 = & \mathbf{u} \cos \psi + \mathbf{v} \sin \psi, \\
 \mathbf{\epsilon}_2 = & - \mathbf{u} \sin \psi + \mathbf{v} \cos \psi.
\end{align}
The tensors for $+$ and $\times$ polarizations can be generated,
\begin{align}
 \mathbf{e}_+ = & \mathbf{\epsilon}_1 \otimes \mathbf{\epsilon}_1 - \mathbf{\epsilon}_2 \otimes \mathbf{\epsilon}_2, \\
 \mathbf{e}_\times = & \mathbf{\epsilon}_1 \otimes \mathbf{\epsilon}_2 + \mathbf{\epsilon}_2 \otimes \mathbf{\epsilon}_1.
\end{align}
Therefore, the GW responded in a laser link from the sender at $\mathbf{p}_s$ to the receiver at $\mathbf{p}_r$ could be expressed as 
\begin{equation}
\begin{aligned}
    h_\mathrm{SSB} (t, \mathbf{p} )  =&  \mathbf{n}_{sr} \cdot \mathbf{e}_+ \cdot \mathbf{n}^T_{sr} \  H_{+}(t, \mathbf{p} ) \\
     & + \mathbf{n}_{sr} \cdot \mathbf{e}_\times \cdot \mathbf{n}^T_{sr} \ H_\times(t, \mathbf{p} ),
\end{aligned}
\end{equation}
where $\mathbf{n}_{sr}$ is the unit vector from sender to receiver, and $H_{+, \times}$ represents the incoming GW signal from a source.
The arrival time $t_r$ with the influence of GW could be obtained from
\begin{equation} \label{eq:reception-time}
    t_r \simeq t_s + \frac{L_{sr}}{c} - \frac{1}{2c} \int_0^{L_{sr}}{h_\mathrm{SSB} \left( t(\lambda), \mathbf{p}(\lambda) \right) d \lambda },
\end{equation}
where $L_{sr}$ is propagation time in the solar systems dynamics obtained from Eqs. \eqref{eq:L_ij}-\eqref{eq:iteration}, and the last term on the right side is the effects of GW. Approximating the wave propagation time as $t(\lambda) \approx t_s + \lambda/c$ and position as $\mathbf{p}(\lambda) = \mathbf{p}_s(t_s) + \lambda \mathbf{n}_{sr}(t_s)$, the GW at a given time point can be expressed as
\begin{equation} \label{eq:H}
    h_\mathrm{SSB} \left[ \mathbf{p}(\lambda), t(\lambda) \right]
    =  h_\mathrm{SSB} \left[ t(\lambda) - \frac{\mathbf{k} \cdot \mathbf{p}(\lambda)}{c} \right].
\end{equation} 
The GW response in relative frequency deviation from sender to receiver could be written as \cite{Katz:2022yqe}
\begin{widetext}
\begin{equation} \label{eq:response-to-gw}
    y_{sr}(t_r) = \frac{1}{2 \left[ 1 - \mathbf{k} \cdot \mathbf{n}_{sr}(t_s) \right]} \left[ h_\mathrm{SSB} \left(t_s - \frac{\mathbf{k} \cdot \mathbf{p}_s (t_s)}{c} \right) - h_\mathrm{SSB} \left(t_r - \frac{\mathbf{k} \cdot \mathbf{p}_r(t_r)}{c} \right) \right].
\end{equation}
With the responded GW in a single link, the time-domain waveform in a TDI channel will be 
\begin{equation}
\begin{aligned}
 h_\mathrm{TDI} (t) = & \sum_{sr, k+}  {y}_{sr} (t + \tau_{k+}) - \sum_{sr, k-}  {y}_{sr} (t + \tau_{k-} ),
 \end{aligned}
\end{equation}
where the first term on the right side is cumulative effects of links from one (adding) beam, and the second term denotes the combination of another beam which will be subtracted from the first term.
The response in the frequency domain could be expressed as
\begin{equation}
    \tilde{y}_{sr}(t_r, f) = \frac{ \tilde{h}_\mathrm{}(f) }{2 \left[ 1 - \mathbf{k} \cdot \mathbf{n}_{sr}(t_s) \right]} \left[ \exp \left( 2 \pi f i \left(t_s - \frac{\mathbf{k} \cdot \mathbf{p}_s (t_s)}{c} \right) \right) - \exp \left( 2 \pi f i \left(t_r - \frac{\mathbf{k} \cdot \mathbf{p}_r(t_r)}{c} \right) \right) \right] .
\label{eq:response-to-gw-fd}
\end{equation}
The frequency-domain waveform of TDI will be
\begin{equation}
\begin{aligned}
 \tilde{h}_\mathrm{TDI} (t, f) = & \sum_{sr, k+}  \tilde{y}_{sr} (t + \tau_{k+}, f) - \sum_{sr, k-}  \tilde{y}_{sr} (t + \tau_{k-}, f).
 \end{aligned}
\end{equation}
The frequency evolution with time is calculated to obtain the instantaneous location of the detector and its response, 
\begin{equation}
t_{lm} (f) = t_c - \frac{1}{2 \pi} \frac{ \mathrm{d} \phi_{lm}(f) }{ \mathrm{d} f}.
\end{equation}
In contrast to Eq. \eqref{eq:response-to-gw}, an approximation is usually applied $\mathbf{p}_s (t_s) \approx \mathbf{p}_s (t_r)$, and the GW in the single link is written as
\begin{equation} \label{eq:response-to-gw2}
    y_{sr, \mathrm{approx}}(t_r) \approx \frac{1}{2 \left[ 1 - \mathbf{k} \cdot \mathbf{n}_{sr}(t_r) \right]} \left[ h_\mathrm{SSB} \left(t_r - \frac{L_{sr}(t_r)}{c} - \frac{\mathbf{k} \cdot \mathbf{p}_s (t_r)}{c} \right) - h_\mathrm{SSB} \left(t_r - \frac{\mathbf{k} \cdot \mathbf{p}_r(t_r)}{c} \right) \right].
\end{equation}
\end{widetext}
For the LISA mission, the velocity of the S/C in SSB coordinates will be $\sim$30 km/s, and the position of sender will deviate from the original location by $\sim$250 km during a transmission time of 8.33 s. Therefore, the approximation in Eq. \eqref{eq:response-to-gw2} may introduce some inaccuracy in the calculation. We choose one source for simulation to examine the deviations between two algorithms given by Eqs. \eqref{eq:response-to-gw} and \eqref{eq:response-to-gw2}. The parameters for waveform simulation are shown in Table. \ref{tab:inj_source_paras}.
 
\begin{table}[tbh]
\caption{\label{tab:inj_source_paras}
The parameters of simulated MBHB source.}
\begin{ruledtabular}
\begin{tabular}{ccc}
 parameter &  value & description  \\
\hline
 $m_1$ & $6 \times 10^4$ $M_\odot$ &  mass of primary BH in source frame \\
 $m_2$ & $1 \times 10^4$ $M_\odot$ & mass of secondary BH in source frame \\
 $d_L$ & 6791.8 Mpc & luminosity distance  \\
 & & (redshift = 1.0 \cite{Planck:2018vyg,Astropy:2013muo}) \\
 $\iota$ & $\pi/3$ rad & inclination angle  \\
 $\lambda$ & 4.603 rad & ecliptic longitude \\
 $\beta$ &  0.314 rad & ecliptic latitude \\
 $\psi$ & 0.55 rad  & polarization angle \\
 $t_c$ & 56.0 day & merger time at SSB center \\
 & & with respect to initial orbit time \\
 $\phi_c$ &  & reference phase \\
 duration  & 25 days  & duration of chirp signal
\end{tabular}
\end{ruledtabular}
\end{table}

\begin{figure}[htb]
\includegraphics[width=0.45\textwidth]{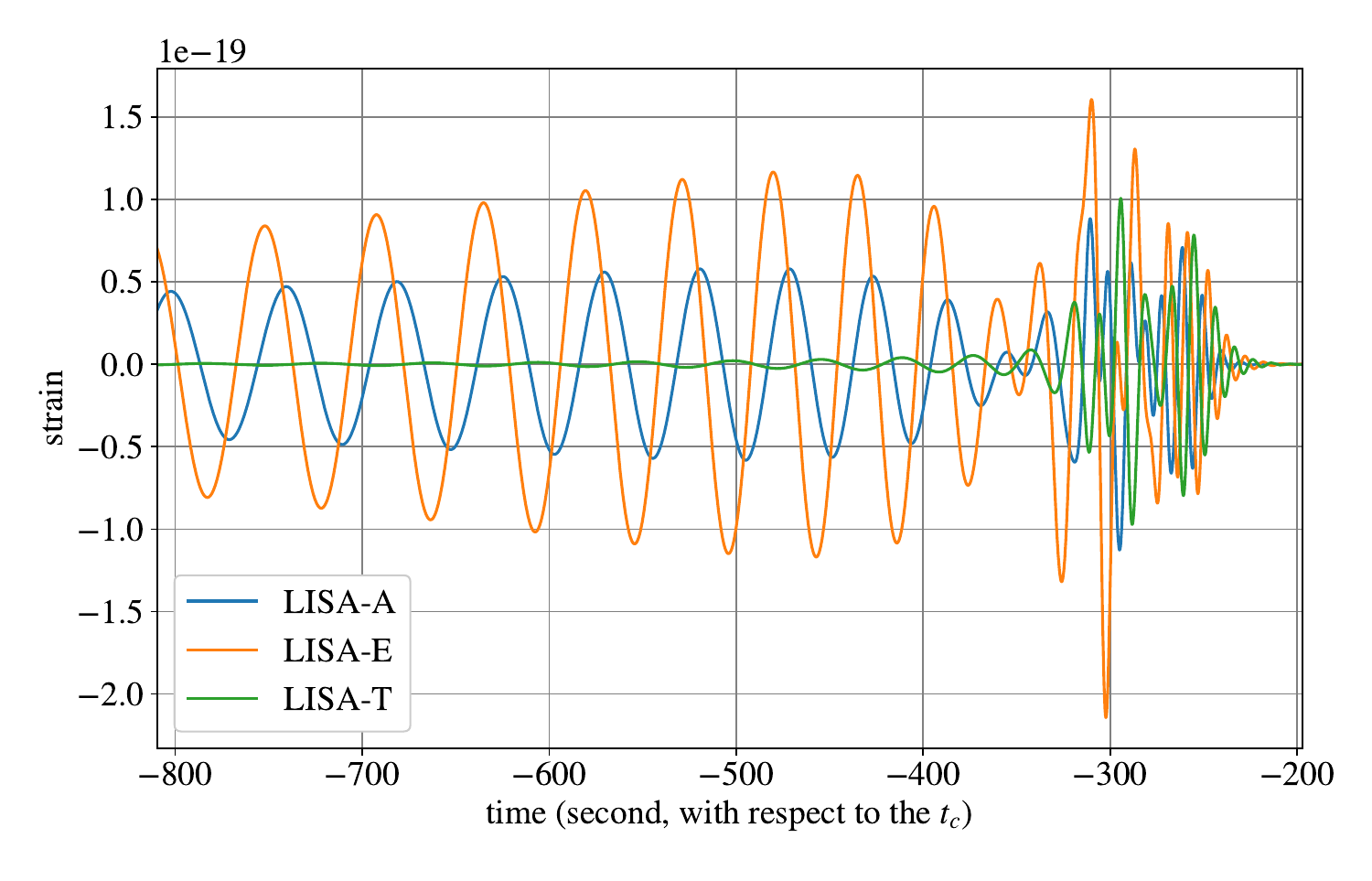} 
\includegraphics[width=0.45\textwidth]{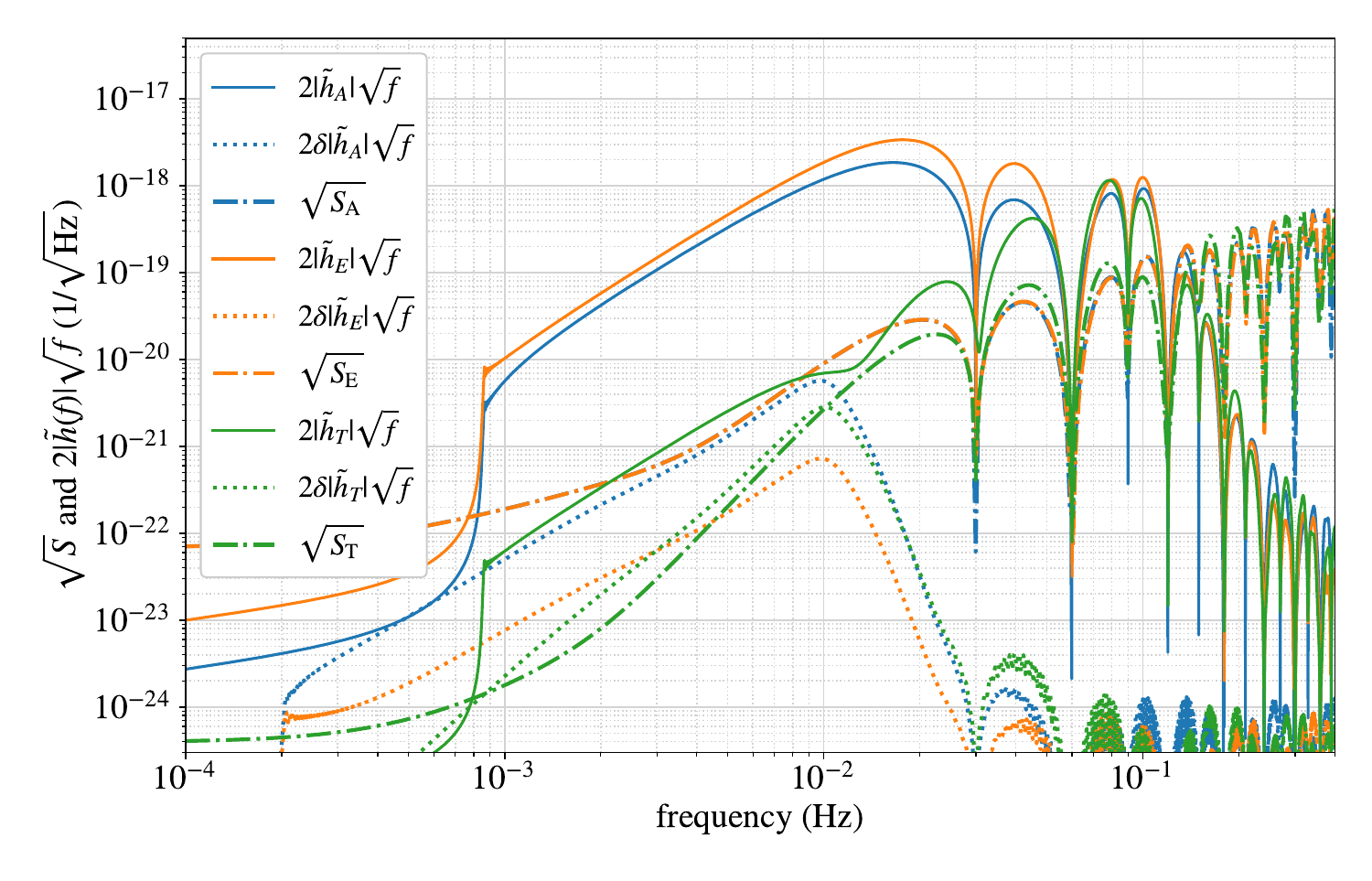}
\includegraphics[width=0.45\textwidth]{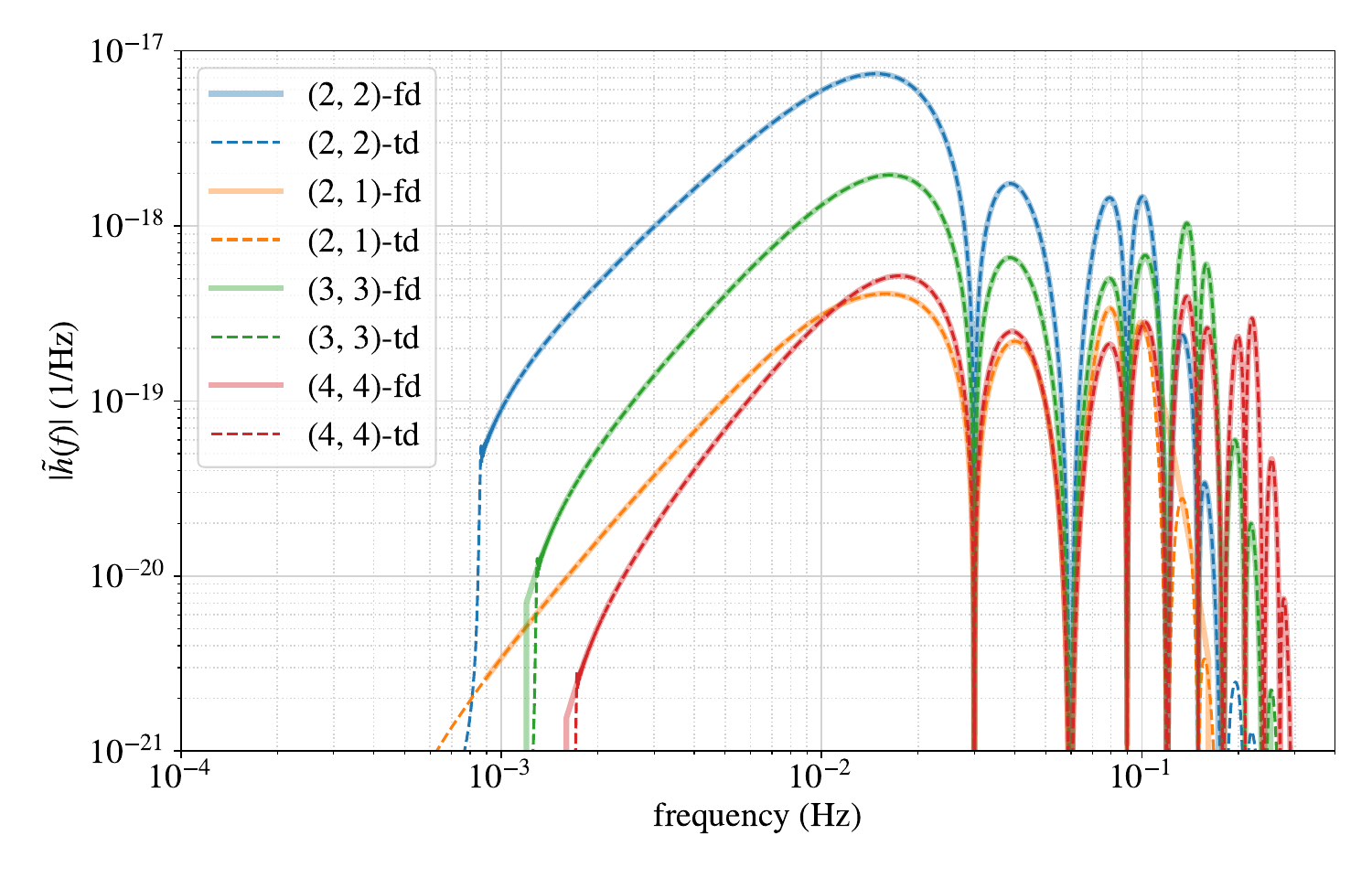} 
\caption{\label{fig:waveform_simulation} The simulated waveforms in optimal channels of second-generation TDI Michelson configuration. The time-domain waveforms are shown in the upper panel. The x-axis is the TDB time with respect to the merger signal arriving at the SSB center, and coalescence reaches the detector earlier than that at SSB center for the current source. The waveforms and waveform differences via Fourier transform of time-domain waveforms are shown in the middle plot, and the differences, $2 \delta|\tilde{h}_\mathrm{TDI}| \sqrt{f}$, are yielded by using the algorithms of Eq. \eqref{eq:response-to-gw} or \eqref{eq:response-to-gw2}. The lower plot shows the multiple harmonic modes $(l, m)$ in the optimal A channel which is generated from time-domain (td) and frequency-domain model (fd).}
\end{figure}

The simulated waveforms of the second-generation TDI Michelson configuration are shown in Fig. \ref{fig:waveform_simulation}. In the upper panel, the time-domain waveforms in three optimal channels are present for the last $\sim$600 before the coalescence. The x-axis is the time in TDB relative to the arrival time of the merger signal at the SSB center. And the coalescence reaches the detector earlier than at the SSB center in current case. As depicted by the waveforms, the amplitudes and phases of waveforms are modulated by the TDI. The amplitudes of waveforms in A and E are significantly suppressed around $t \simeq -350$ s, corresponding to their lowest characteristic frequency, $f_c = 1/(4L) \simeq 0.03$ Hz. For higher frequencies, the wavelength of GW is shorter than $4L$, causing the mixed phase to cross one cycle in TDI. The different amplitudes of A and E also verify that two observables have different antenna patterns \cite{Wang:2020fwa}, despite two channels having identical averaged-sensitivities. The waveform in the T channel is noticeably canceled in low frequencies, and the amplitude is amplified with the increase of frequencies. 
 
In the middle plot of Fig. \ref{fig:waveform_simulation}, Fourier transforms are applied to the time-domain waveforms, as well as the differences between two waveforms ($ \delta h_\mathrm{TDI} = h_\mathrm{TDI} - h_\mathrm{TDI, approx}$), generated from Eqs \eqref{eq:response-to-gw} and \eqref{eq:response-to-gw2}. These differences are depicted by the dotted lines, and the noise ASDs of TDI channels are also shown for comparison. The ASDs of A (blue dashed-dotted line) and E (orange dashed-dotted line) are equal and overlapped, and the T channel exhibits lower than the other two channels at lower frequencies and becomes comparable at high frequencies. 
For current source, the differences between two algorithms vary among three channels. In the T channel, the discrepancy could exceed its noise level within the frequency range of [1 mHz, 10 mHz]. The residual waveform in the A channel may be comparable to the noise level at certain frequencies, whereas the waveform difference in the E channel is one order of magnitude lower than its noise. The divergence levels of GW signals may vary with different sources. On the other hand, the waveforms in the frequency domain clearly show that GW signals are significantly suppressed by TDI around their characteristic frequencies ($f_c =\frac{n}{4L} $ for $n=1, 2, 3...$). 

In the lower panel of Fig. \ref{fig:waveform_simulation}, in addition to the predominant $(l=2, m=2)$ harmonic mode of the GW, extra modes of waveform are simulated and plotted for the A channel. The dashed lines indicate the waveforms from time-domain simulation, and the solid lines represent the waveforms obtained by using the frequency-domain model. As shown by the curves, the waveforms exhibit good agreements in terms of amplitudes. However, the waveforms from time-domain and frequency-domain approximants are best fitted at different phases/times. As a result, their match decreases when matched filtering is implemented with integrated multiple harmonics. As a compromise, only the dominant $(l=2, m=2)$ mode is injected into the noise data to perform the analysis.

\section{Analysis} \label{sec:analysis}

The simulated data is synthesized by injecting the waveforms into the noise data. Initially, the ordinary data streams of the first-generation TDI observables (X, Y, Z) and the second-generation observables (X1, Y1, Z1) are generated. Then, the optimal data streams are composed using Eq. \eqref{eq:abc2AET}. The analysis strategy in current study comprises two steps, the first step involves matched filtering for signal identification, and the second step focuses on parameter inference for the binary.

\subsection{Identification of chirp signal}

The fiducial strategy for identifying GW chirp signals involves filtering the data with a pre-built template bank \cite[and references therein]{Allen:2005fk,Usman:2015kfa,Nitz:2017svb,Weaving:2023fji}. The template bank should adequately cover the targeting parameter space with a required density \cite[and references therein]{Babak:2012zx,Brown:2012qf,DalCanton:2014hxh,Sakon:2022ibh}. Machine learning techniques have been applied to identify the chirp signals in advanced LIGO/Virgo data \cite{Gabbard:2017lja,Schafer:2022dxv}. For ground-based GW searches, chirp signals typically last seconds to dozens of seconds in the sensitive band of detectors, and the motion of detector with the Earth could be ignored. The template bank is built to cover the intrinsic parameters of the binary sources, such as masses and spins. 
However, for space-borne GW detectors, chirp signals could persist for days to months in their sensitive band, and the motion of interferometer could alter the response function to GWs. Moreover, TDI introduces the additional modulation into the observed signals. Therefore, the template bank may need to include extrinsic parameters to sufficiently match the observed signals. \citet{Weaving:2023fji} generated a template bank searched for MBHB in the LISA mock data, and verified the feasibility of the signal identification strategy. 
Given that the template bank searches are computing expensive, we only introduce the matched filtering algorithm here.

The matched filter is an optimal algorithm for identifying a GW signal with a known waveform. The output of the filtering can be expressed as \cite{Allen:2005fk}
\begin{equation}
    z(t_0) = 4 \Re \int^{\infty}_0 \frac{ \tilde{d} (f) \tilde{h}^{\ast}_\mathrm{TDI} (f) }{ S_\mathrm{TDI} (f) } e^{2 \pi i f t_0} d f,
\end{equation}
where $\tilde{d}$ is the data in the frequency domain, $\tilde{h}_\mathrm{TDI}$ is a waveform template, $S_n$ is the noise PSD of the data stream, and $t_0$ is a reference time which refers to coalescence time here. A normalization factor can be obtained for the template, 
\begin{equation}
    \sigma^2_{m} = 4 \int^{\infty}_0 \frac{ | \tilde{h}_\mathrm{TDI} (f) |^2 }{S_\mathrm{TDI} (f)} d f.
\end{equation}
Then, the signal-to-noise ratio (SNR) of the matched filtering is given by
\begin{equation}
    \rho(t) = \frac{ | z(t) | }{ \sigma_m }.
\end{equation}

\begin{figure}[htb]
\includegraphics[width=0.48\textwidth]{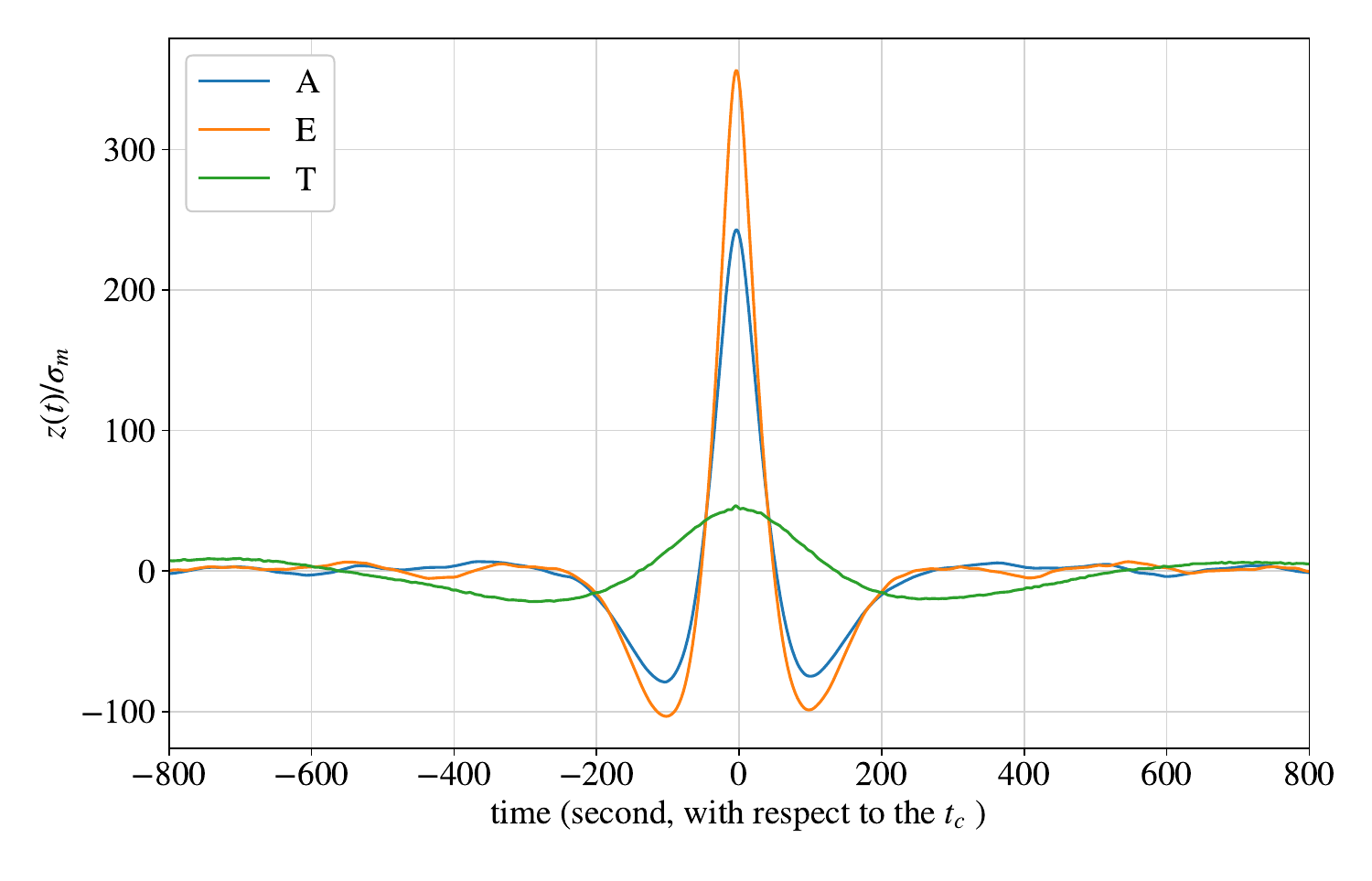}
\includegraphics[width=0.48\textwidth]{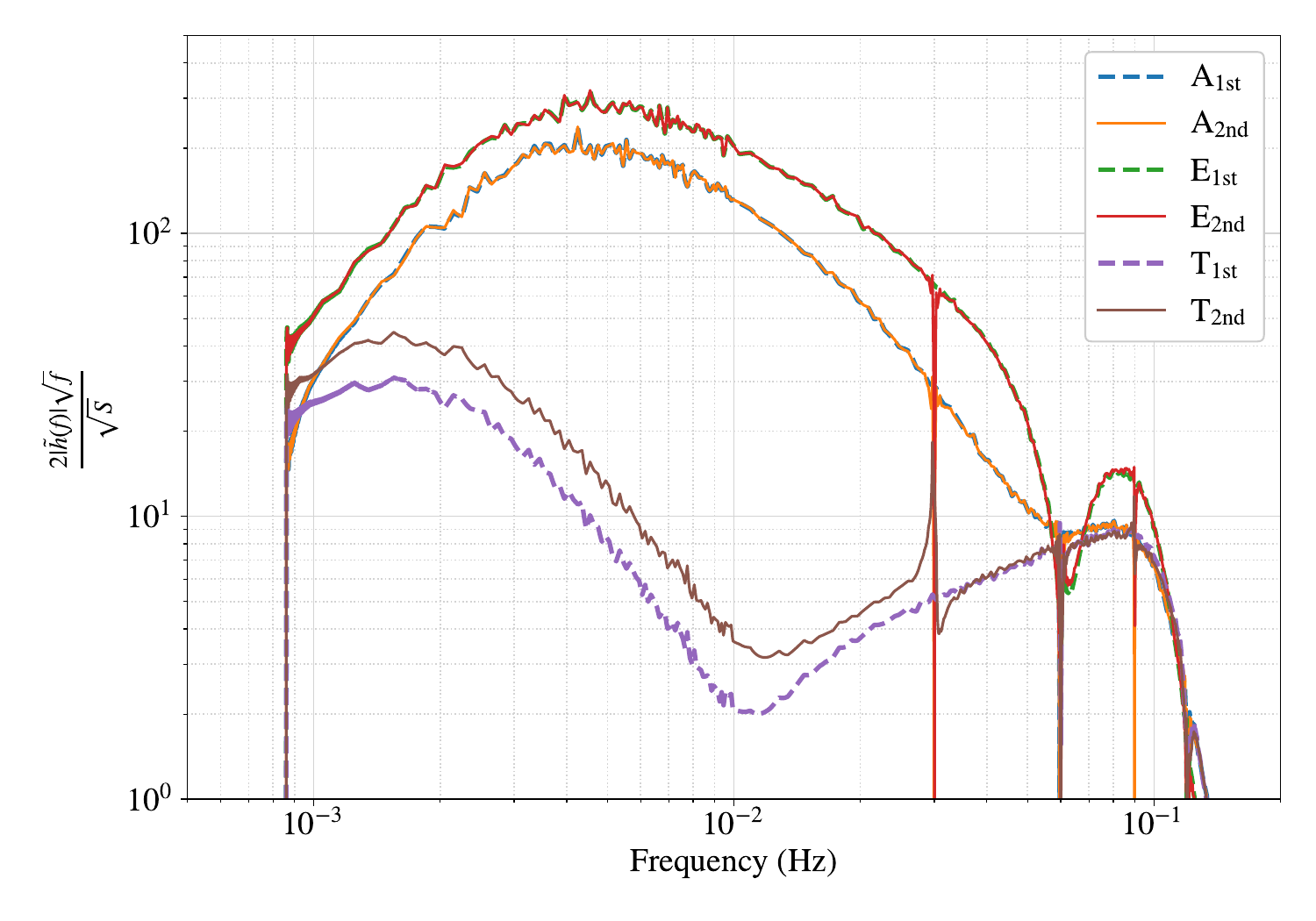}
\caption{\label{fig:snr_matched_filter} The matched filter outputs for optimal channels from second-generation TDI Michelson (upper) and the ratio $\frac{2 |\tilde{h}| \sqrt{f}}{\sqrt{S}}$ for optimal channels from both the first-generation (1st) and second-generation (2nd) TDI Michelson configurations (lower).}
\end{figure}

Using a frequency-domain waveforms modeled with fitted parameters in Section \ref{subsec:waveform}, a matched filter is implemented to the optimal channels of the second-generation TDI Michelson, and the results are presented in the upper panel of Fig. \ref{fig:snr_matched_filter}. Among the three optimal channels, the E channel exhibits the highest SNR ($\rho_E=356$), followed by the A channel with a lower SNR ($\rho_A=243$). The T channel shows a significantly lower SNR ($\rho_T=45$) compared to the other two channels. Additionally, the SNR profile of the T channel is broader than that of other two channels, and this should be due to that its SNR is mainly contributed from lower frequencies. 

An optimal SNR is defined for the template fully matching the signal in data,
\begin{equation}
\begin{aligned}
    \rho^2_\mathrm{opt} = & 4 \int^{\infty}_0 \frac{ | \tilde{h}_\mathrm{TDI} (f) |^2 }{S_\mathrm{TDI} (f)} d f \\
     = & \int^{\infty}_0 \left( \frac{  2 | \tilde{h}_\mathrm{TDI} (f) | \sqrt{f} }{ \sqrt{S_\mathrm{TDI} (f) } } \right)^2 d \ln f.
\end{aligned}
\end{equation}
The ratio $\frac{2 |\tilde{h}| \sqrt{f}}{\sqrt{S}}$ between the waveform and noise ASD partially reflect the SNR contribution across a log-scaled frequency range. The ratios for Michelson configuration from both first-generation and second-generation TDI are depicted in the lower panel of Fig. \ref{fig:snr_matched_filter}. As expected, the E channels exhibit the highest values among three observables, and the A channel is slightly lower than E for this instance. The A/E observables from both first-generation and second-generation TDI configurations are (nearly) identical, except at characteristic frequencies. Their largest SNR contributions are expected to occur in the frequency band around 5 mHz. However, noticeable differences are observed in the T channels between two different generations. 
As we have previously analyzed in \cite{Wang:2020fwa,Wang:2024b}, benefiting from unequal arm interferometry, the T channels from Michelson TDI are more sensitive compared to the equal-arm case. The T channel may contain GW signals resulting from imperfect cancellation of the three ordinary channels (X, Y, Z) or (X1, Y1, Z1). The second-generation observables process more recurrent links than the first-generation, amplifying the effect of unequal arm length. Consequently, the SNR in the T channel of the second-generation is higher than that of the first-generation. On the other side, in these observables, especially for the second-generation, the ratios exhibit discontinuities at their characteristic frequencies, caused by the overestimated PSDs at these frequencies due to averaging.

\subsection{Parameter inference}

Source parameters are inferred from the observed data by using Bayesian analysis. The posteriors of parameters are calculated from
\begin{equation}
p( \vec{\theta} | d )  = \frac{ \mathcal{L}(d | \vec{ \theta } ) p(\vec{\theta}) }{ p(d) },
\end{equation}
where  $p(\vec{\theta})$ is prior probabilities of parameters. $\mathcal{L}(d|\vec{\theta} )$ is the likelihood function of a set of parameters \cite{Romano:2016dpx},
\begin{equation}
\begin{split}
\ln \mathcal{L} (d|\vec{\theta} ) = & \sum_{f_i} \left[ 
-\frac{1}{2} (\tilde{\mathbf{d}} - \tilde{\mathbf{h}} ( \vec{\theta}) )^T \mathbf{C}^{-1}  (\tilde{\mathbf{d}} - \tilde{\mathbf{h}}(\vec{\theta}) )^\ast \right. \\
 &  \left.  -\frac{1}{2} \ln \left( \det{ 2 \pi \mathbf{C} } \right)
   \right],
\end{split}
\end{equation}
and $\mathbf{C}$ is the correlation matrix of noises in selected TDI channels,
\begin{equation}
 \mathbf{C} = \frac{T_\mathrm{obs}}{4}
 \begin{bmatrix}
S_{xx} & S_{xy} & S_{xz} \\
S_{yx} & S_{yy} & S_{yz} \\
S_{zx} & S_{zy} & S_{zz}
\end{bmatrix},
\end{equation}
where $T_\mathrm{obs}$ is the observation time which is 30 days in this investigation, $S_{xx}$/$S_{xy}$ are PSD or cross-spectral density of data streams which estimated by using algorithm detailed in \cite{Allen:2005fk}. 
For networks with multiple detectors, the joint log-likelihood is given by
\begin{equation}
\ln \mathcal{L}_\mathrm{joint} =  \sum_\mathrm{det} \ln \mathcal{L}_\mathrm{det} (d|\vec{\theta} ) 
\end{equation}

The prior setups are shown in Table \ref{tab:priors}. Instead of inferring individual component masses ($m_1, m_2$), the redshifted chirp mass $\mathcal{M}_c = (m_1 m_2)^{3/5} / (m_1 + m_2)^{1/5}$ and mass ratio $q = m_2/m_1$ are estimated. The prior of chirp mass is adjustable for different mass binaries, and the prior of mass ratio is set to be a range of [0.1, 0.2]. The prior ranges of these two parameters are set to be relatively smaller to reduce computing time, and it will not affect the final results. The luminosity distance $d_L$ uses a distribution in the comoving volume by using the cosmological parameters from Planck 2018 \cite{Planck:2018vyg,Astropy:2013muo}. The inclination $\iota$ follows a sinusoidal distribution, the sky location ($\lambda, \beta$) is uniformly distributed in a spherical surface, the reference phase $\phi_c$ is evenly distributed in $[0, 2 \pi]$, and the coalescence time $t_c$ is uniformly sampled within $\pm$180 seconds with respect to the injected time. The Nested sampler $\mathsf{MultiNest}$ \cite{Feroz:2008xx,Buchner:2014nha} is employed to perform the parameter inference.

\begin{table}[tbh]
\caption{\label{tab:priors}
Priors for parameter estimation. $\mathcal{M}_c = (m_1 m_2)^{3/5} / (m_1 + m_2)^{1/5}$ is redshifted chirp mass, $q=m_2/m_1$ is the mass ratio, $d_L$ is the luminosity distance of source, $\iota$ is inclination angle between the orbital angular momentum and line of sight to observers, $(\lambda, \beta)$ are the ecliptic longitude and latitude of source, $\psi$ is polarization angle, $t_c$ is arrival time of coalescence at SSB center, and $\psi_c$ is reference phase.}
\begin{ruledtabular}
\begin{tabular}{cccc}
Variable & Prior &  range & unit  \\
\hline
$\mathcal{M}_c$ & uniform &  & $M_\odot$  \\
$q$ & uniform & [0.1, 0.2] &  \\
$d_L$ & Comoving & [5000, 8000] & Mpc  \\
$\iota$ & $\sin$ & [0, $\pi$] & rad  \\
$\lambda$ & uniform & [0, $2 \pi$] & rad  \\
$\beta$ & $\cos$ & [$ -\frac{\pi}{2},  \frac{\pi}{2}$] & rad  \\
$\psi$ & uniform & [0, $\pi$] & rad  \\
$t_c$ & uniform & [$t_c$-180, $t_c$+180] & second  \\
$\phi_c$ & uniform & [0, $2 \pi$] & rad  \\
\end{tabular}
\end{ruledtabular}
\end{table}

\begin{figure*}[htb]
\includegraphics[width=0.99\textwidth]{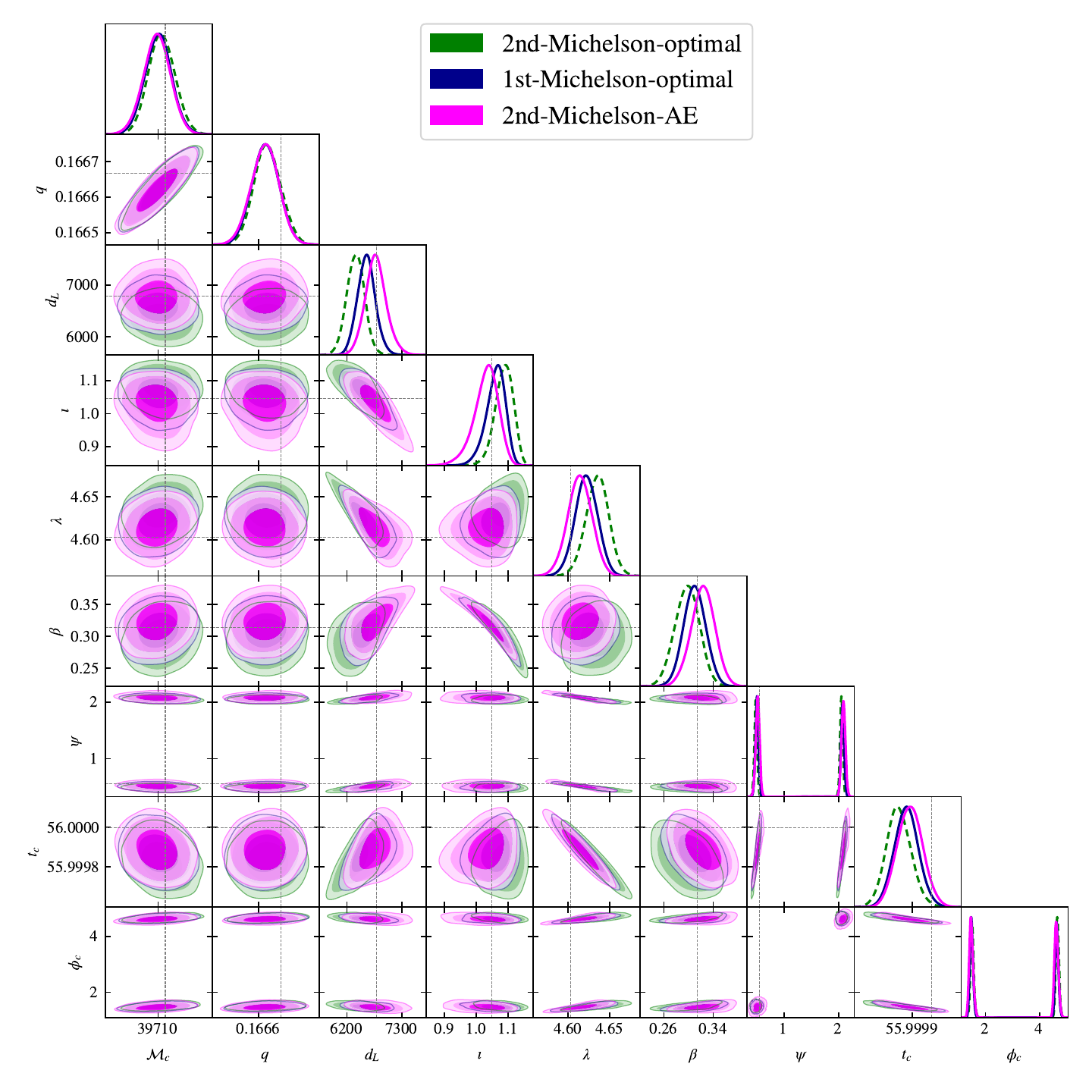}
\caption{\label{fig:corner_m1_6e4} The distributions of parameters inferred from three TDI datasets. The green colors indicates the result from second-generation TDI Michelson optimal channels (A, E, T), the blue color shows the result of first-generation TDI Michelson optimal channels, and the magenta is the results of second-generation TDI Michelson optimal science observables (A, E). Three gradients indicate the $1\sigma$, $2\sigma$ and $3\sigma$ regions from darker to lighter.}
\end{figure*}

Three cases are executed for the first comparison with different TDI datasets: 1) three optimal channels (A, E, T)  of first-generation Michelson, 2) three optimal channels from second-generation Michelson, and 3) two science channels from second-generation Michelson (A, E). The inferred distributions of parameters are shown in Fig. \ref{fig:corner_m1_6e4}. The parameters inferred from three scenarios are moderately different. The distributions of the mass parameters are almost identical, while the distributions for extrinsic parameters vary with datasets. As aforementioned, the detectability of the T channel is sensitive to arm differences during the orbital motion, and the time-varying information could be averaged and discarded in the frequency-domain analysis. As a result, accurately inferring the signal in the T channel could be tricky or otherwise cause bias in estimation. As shown in the result, since the T channel from second-generation Michelson accumulates more SNR than it does in first-generation, the distributions of extrinsic parameters inferred from second-generation channels diverge from the injected values more than the first-generation, especially for the luminosity distance $d_L$ and longitude $\lambda$. After removing the T data stream, two science channels from the second-generation can constrain all parameters in sensible regions.
As we also can see from the corner plot, the $\mathcal{M}_c$ and $q$ are correlated as expected. The inclination $\iota$ not only degenerates with the distance $d_L$ but also the ecliptic latitude $\beta$, and the longitude is correlated with arrival time $t_c$. Both polarization angle $\psi$ and reference phase $\phi_c$ are inferred with bimodal distributions. 

\begin{figure*}[htb]
\includegraphics[width=0.99\textwidth]{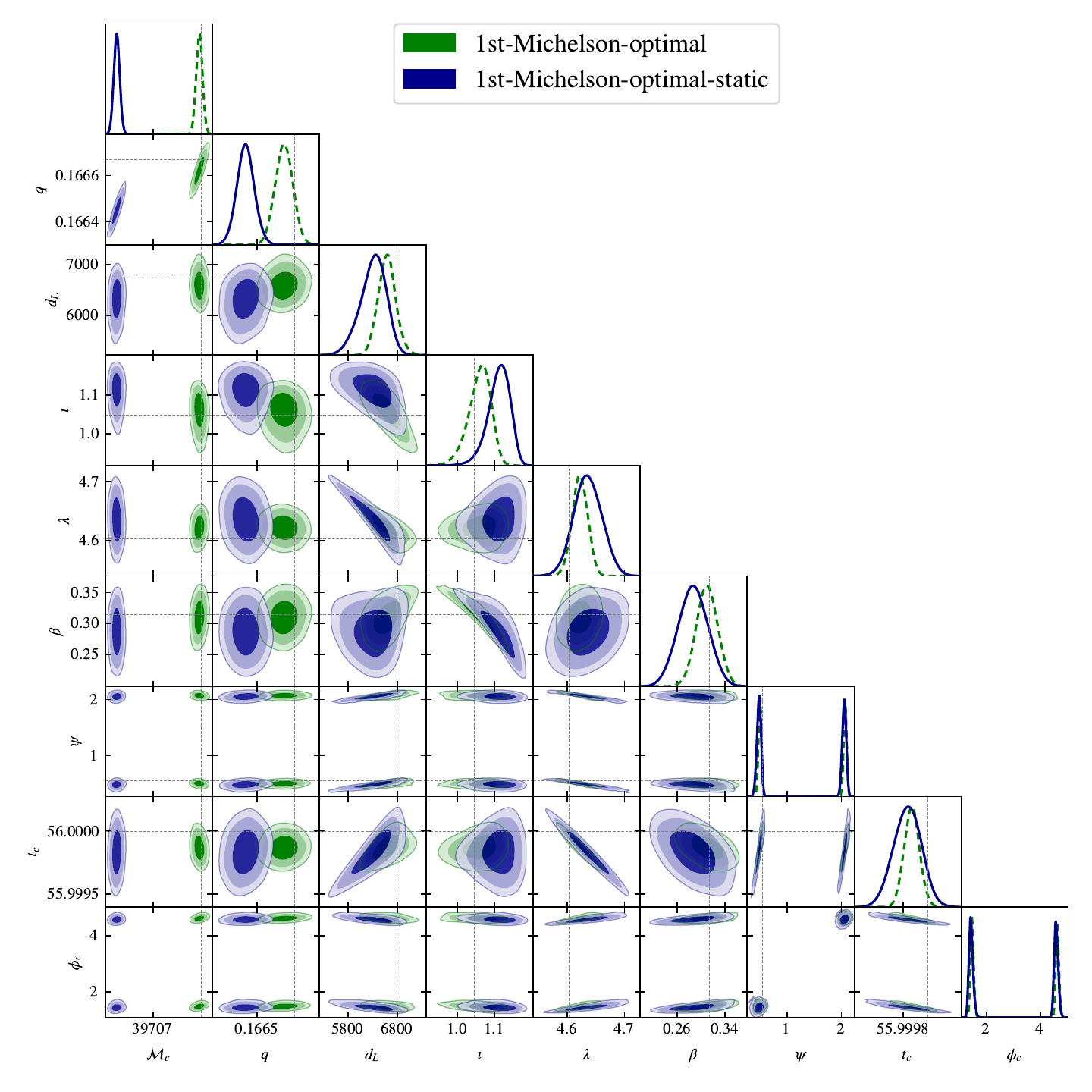}
\caption{\label{fig:corner_m1_6e4_vs_static} The parameter inference results from orbital dynamic model and orbital static model. The green colors show the result from the first-generation TDI Michelson optimal channels considering detector's orbital motion, the blue colors indicate the results of same TDI observable with assumption of a static detector at coalescence time for 30 days.}
\end{figure*}

In another case, for a LISA-like mission, the orbital motion of interferometer will modulate the amplitude and phase of waveform. Hence, the effects of orbit dynamics should be taken into account when modeling TDI waveforms in the frequency-domain, especially for a long-duration signal. To access this impact, an inference with static constellation is performed using the first-generation Michelson optimal channels. The results are presented in Fig. \ref{fig:corner_m1_6e4_vs_static}. Compared to the dynamic model, the inferred values from the static model diverge from the true values, particularly for the intrinsic parameters, chirp mass $\mathcal{M}_c$ and mass ratio $q$, while the extrinsic parameters are still within the $3\sigma$ credible regions. Furthermore, the extrinsic parameters estimated from dynamic model exhibit better constraints with smaller uncertainties. One reason for this could be the better matched waveforms, and another reason could be that the motion of detector helps to resolve the location of source by demodulated response function.

\begin{figure*}[htb]
\includegraphics[width=0.99\textwidth]{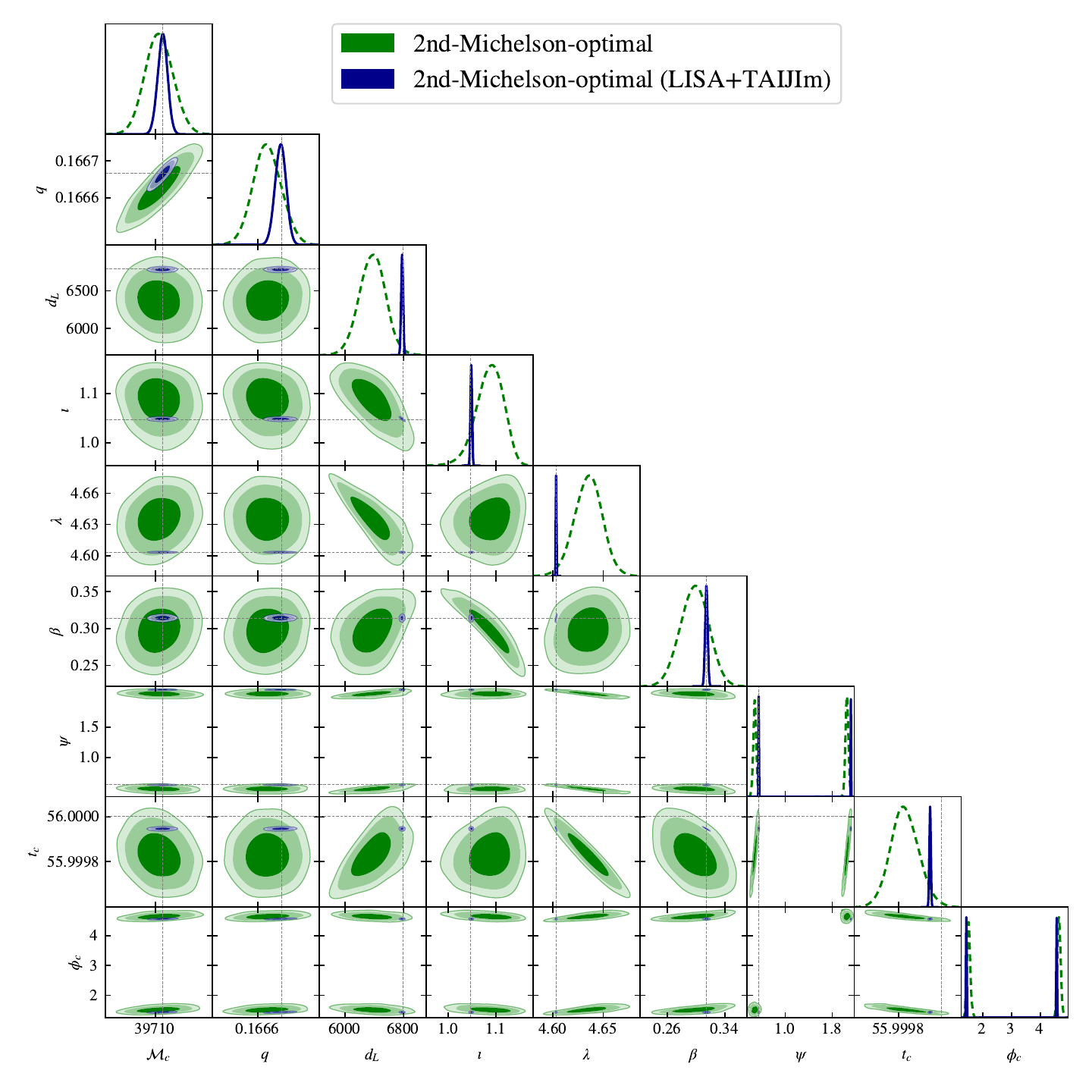}
\caption{\label{fig:corner_m1_6e4_joint} The parameter inference results from single LISA mission and LISA-TAIJIm network. The green curves show the results from the second-generation TDI data streams of LISA, the blue colors indicate the result from LISA-TAIJIm joint analysis.}
\end{figure*}

In principle, simulating and analyzing a variety of sources would be necessary to validate the analysis performance. However, this process would be computationally costly. To assess the analysis performance more rigorously, a more challenging case could be utilized to examine whether the parameters can be accurately inferred. As proposed in \cite{Wang:2021uih}, the inclusion of the TAIJIm observatory in the network with LISA can significantly promote the determinations of parameters for MBHBs. By jointly analyzing simulated data for TAIJIm along with LISA, the result, shown by the blue regions in Fig. \ref{fig:corner_m1_6e4_joint}, indicates that all parameters can be inferred accurately with much smaller uncertainties. This suggests the correctness and robustness of analysis. Additionally, the joint data analysis also mitigates the inaccurate influence from the Michelson T channels and produces sensible distributions.

\section{Conclusion and discussion} \label{sec:conclusions}

In this work, we introduce a generic framework designed for simulating and analyzing TDI data. Our objectives include investigating key factors influencing TDI implementations, developing algorithms for GW analysis, and exploring the potential of alternative TDI observables. By utilizing numerical orbit simulations, the framework is capable of generating time-domain data comprising both noises and GW signals in TDI channels. Additionally, the current version of the framework incorporates parameter inference by matched filtering the modeled GW signal for MBHB. We specifically focus on simulating and analyzing a chirping GW signal using data streams from both first-generation and second-generation TDI Michelson configurations. The results validate the effectiveness of the framework and reveal the potential drawbacks of the fiducial Michelson configuration under scenarios with dynamic unequal arm lengths. Furthermore, we propose an alternative TDI configuration to mitigate the impacts of unequal arm and null frequencies in a companion paper \cite{Wang:2024b}.

LISA is chosen as the typical mission for demonstrating the simulation and analysis in this study, and the framework is also applicable to other space-based GW missions employing TDI. The validation of these processes are carried out by selecting MBHBs, which are the primary targets of LISA. In the data analysis, harmonic modes of GWs beyond the $(l=2, m=2)$ are omitted, although incorporating additional modes could improve parameter inference. In addition, the Doppler effect resulting from detector motion is disregarded, as the modulation is expected to be insignificant for a one-month chirp signal at current SNR levels. However, this modulation could become significant and should be considered for less massive binaries with longer durations, such as stellar-mass compact binaries. Extended analysis incorporating greater completeness will be pursued in future work. 

A challenge in GW observations in the milli-Hz band is the simultaneous detection of various sources, both of the same and different types, leading to heavily overlapped signals. Additionally, instrumental noises may vary over time, and observed data may not be continuous. As the framework evolves, these effects will be integrated into the simulation, warranting comprehensive investigations into their impacts on GW analysis. Moreover, networks with multiple detectors show promise for boosting GW detections in space \cite{Cai:2023ywp}, and the joint analysis could offer significant benefits, as tentatively explored in this study. Relevant simulations and analyses will be conducted in future work.

\begin{acknowledgments}

GW thanks Alex Nitz, Runqiu Liu, and Minghui Du for helpful communications and discussions.
GW was supported by the National Key R\&D Program of China under Grant No. 2021YFC2201903, and NSFC No. 12003059. This work made use of the High Performance Computing Resource in the Core Facility for Advanced Research Computing at Shanghai Astronomical Observatory.
This work are performed by using the python packages $\mathsf{numpy}$ \cite{harris2020array}, $\mathsf{scipy}$ \cite{2020SciPy-NMeth}, $\mathsf{pandas}$ \cite{pandas}, PyCBC \cite{Usman:2015kfa,Biwer:2018osg}, LALSuite \cite{lalsuite,swiglal}, $\mathsf{astropy}$ \cite{Astropy:2013muo}, $\mathsf{BILBY}$ \cite{Ashton:2018jfp}, $\mathsf{MultiNest}$ \cite{Feroz:2008xx} and $\mathsf{PyMultiNest}$ \cite{Buchner:2014nha}, and the plots are make by utilizing $\mathsf{matplotlib}$ \cite{Hunter:2007ouj}, $\mathsf{GetDist}$ \cite{Lewis:2019xzd} and $\mathsf{Component Library}$ \cite{ComponentLibrary}.

\end{acknowledgments}

\nocite{*}
\bibliography{apsref}

\end{document}